\documentclass[twocolumn,preprintnumbers]{aastex631}

\usepackage{amsmath}

\shorttitle{}
\shortauthors{}

\graphicspath{{./}{figures/}}
\newcommand*\df{\mathop{}\!\mathrm{d}}

\newcommand{\n}{\nonumber}

\def	\m		{\,{\rm m}}

\begin{document}

\title{Erosion of Icy Interstellar Objects by Cosmic Rays and Implications for `Oumuamua}

\author[0000-0002-5611-095X]{Vo Hong Minh Phan}
\affil{Institute for Theoretical Particle Physics and Cosmology (TTK), RWTH Aachen University, 52056 Aachen, Germany; \href{mailto:vhmphan@physik.rwth-aachen.de}{vhmphan@physik.rwth-aachen.de}}

\author[0000-0003-2017-0982]{Thiem Hoang}
\affil{Korea Astronomy and Space Science Institute, Daejeon 34055, Republic of Korea; \href{mailto:thiemhoang@kasi.re.kr}{thiemhoang@kasi.re.kr}}
\affil{Korea University of Science and Technology, Daejeon 34113, Republic of Korea}

\author[0000-0003-4330-287X]{Abraham Loeb}
\affil{Astronomy Department, Harvard University, 60 Garden Street, Cambridge, MA, USA; \href{mailto:aloeb@cfa.harvard.edu}{aloeb@cfa.harvard.edu}}

\reportnum{TTK-21-35}

\begin{abstract}
We study the destruction and modification of icy interstellar objects by cosmic rays and gas collisions. Using the cosmic-ray flux measured in the local interstellar medium as well as inferred from gamma-ray observations at the different galactocentric radii, we find that cosmic-ray erosion is significant for interstellar objects made of common types of ices. Interestingly, cosmic-ray heating might destroy icy interstellar objects very efficiently such that the initial size of an N$_2$ fragment as suggested by \citet{jackson2021} to explain the composition of `Oumuamua should be at least 0.5 km in size in order to survive the journey of about 0.5 Gyr in the ISM and might be even larger if it originated from a region with an enhanced cosmic-ray flux. This implies an initial N$_2$ mass that is at least an order of magnitude larger than the final value, exacerbating the N$_2$ mass budget deficiency for explaining `Oumuamua. The erosion time due to cosmic-ray heating and gas collisions also allows us to set approximate limits on the initial size for other types of icy interstellar objects, e.g. composed of CO, CO$_2$, or CH$_4$. For a given initial size, we constrain the maximum distance to the birth site for interstellar objects with different speeds. We also find that cosmic-ray and gas heating could entirely modify the ice structure before destroying interstellar objects.
\end{abstract}


\section{Introduction}
The first interstellar object (ISO), 1I/2017 U1 (`Oumuamua) has provoked many discussions ever since its discovery by Pan-STARRS \citep{bacci2017}. The light-curve modeling of \citealt{jewitt2017} suggests that `Oumuamua has an elongated shape with an extreme axial ratio of $\gtrsim 5:1$ \citep[see also][]{gaidos2017,fraser2018,mashchenko2019}. It has been proposed that the bizarre shape might be due to ablation induced by interstellar dust \citep{domokos2017} or rotational disruption of the parent object by mechanical torques \citep{hoang2018}. More recent studies involve also planetesimal collisions \citep{sugiura2019} or tidal fragmentation of the original body close to a dwarf star \citep{zhang2020}.   

Another enduring mystery of `Oumuamua is the non-graviational acceleration, which is closely related to the discussion about its composition \citep{micheli2018}. It is believed that the acceleration excess could be explained by the outgassing of volatiles typically observed for comets. Deep observations the Spitzer space telescope \citep{trilling2018} and Gemini North telescope \citep{drahus2018}, however, found no cometary activity of carbon-based molecules. \cite{bialy2018} proposed the scenario where `Oumuamua is accelerated by radiation pressure as a thin lightsail. 
The acceleration anomaly could also be resolved by assuming a porous object \citep{moro-martin2019} or an icy object of unusual composition \citep{fitzsimmons2018}. In fact, \cite{fuglistaler2018} and \cite{segliman2020} have argued that `Oumuamua is made of H$_2$ ice. However, the level of outgassing required for the acceleration excess might result in a rapid change of `Oumuamua's rotation period which has not been observed. More importantly, \citet{hoang2020} has shown that the sublimation rate of H$_2$ ice might be too severe such that a multi-km H$_2$ object might be completely destroyed within less than $10^8$ years.  

Recent studies by \citet{jackson2021} have proposed that 'Oumuamua is a fragment of N$_2$ ice since an object of this type is more likely to survive the interstellar journey owing to a much lower sublimation rate. The authors have examined also other destruction mechanisms and found that the one due to Galactic cosmic rays (CRs) turns out to be most important. In fact, \citet{jackson2021} have provided the erosion rate by CRs to be about $6$ to $60$ m/Gyr depending on their assumptions about the CR density. This corresponds to an erosion time of approximately 2 to 20 Gyr for an N$_2$ fragment of size 100 meters. We argue in the following that CR impulsive spot heating might result in an erosion time of about 1 to 2 orders of magnitude smaller than the previous estimate. As a result, the initial size of `Oumuamua should be larger than roughly 0.5 km if it is made of N$_2$ and has been travelling in the interstellar medium (ISM) for the last 0.5 Gyr. This seem to pose even more serious challenges to the large mass budget and the high formation efficiency required to explain `Oumuamua as an N$_2$ iceberg of an exo-Pluto surface \citep{desch2021,levine2021,siraj2021}.  

We investigate also the modification of the ice structure for these objects due to CR and gas heating. The results suggest that even if ISOs are initially created with the crystal ice structure, they should be entirely modified before being completely destroyed in their ISM journey.

The paper is organized as follows. In Section \ref{sec:destruction}, we derive the destruction rate for CR heating and gas collisions that are applicable for ISOs made of different types of ices including N$_2$, CO, CO$_2$, or CH$_4$. This destruction rate allows us to derive both the evolution of the object's radius in time and the erosion time in the ISM. We then adopt these results in Section \ref{sec:modification} to study the ice modification for ISOs. In Section \ref{sec:implications}, we provide some numerical estimates for the erosion and ice modification times which are then compared with the travel time for ISOs of different sizes to provide insights into the composition and the origin of `Oumuamua. Our main findings are summarized in Section \ref{sec:conclusions}.

\section{Destruction of Icy Interstellar Objects in the Interstellar Medium}
\label{sec:destruction}

There are many different destruction mechanisms for icy ISOs in the ISM. It has been shown that thermal sublimation is efficient only for ices with low binding energy, e.g. H$_2$ and Ne ices \citep{hoang2020,jackson2021}. In fact, for other types of ices, spot heating by CRs and interstellar gas is more important and shall be addressed in details below. We note that erosion due to photodesorption, CR sputtering, or collisional heating by dust grains are also discussed in the literature but they are, in general, less efficient than the processes considered in this work \citep{hoang2020}.  

\subsection{Cosmic-Ray Impulsive Spot Heating}
We will estimate the rate of destruction of icy ISOs due to impulsive spot heating by CRs following the method as presented in \cite{hoang2020}. Since the stopping length of 10 GeV CRs is either much larger than or of order the ISO size considered, we shall limit ourselves to CRs of energies up to $E_{\rm max}=10$ GeV and assume that these particles deposit all their energies into heat for transient evaporation of the ISO ice molecules. The contribution CR electrons and heavy nuclei will be neglected as their deposited energies are subdominant in comparison to CR protons within the energy range considered. We shall first approximate the ISO to a spherical shape of effective radius $R$. The effect of a non-spherical shape will be discussed in Section \ref{sec:implications}. After hitting the ISO, CRs are expected to form tracks of high enough temperature to impulsively evaporate all the ice molecules along these tracks. The evaporated volume per unit of time due to CRs could be estimated from the conservation of energy \citep{leger1985,white1996,hoang2020}
\begin{eqnarray}
n_{\rm ice} E_b\left(\frac{\df V}{\df t}\right)_{\rm CR}
%=4\pi R^2\int^{E_{\rm max}}_{E_{\rm min}} 4\pi j(E) E \df E
=4\pi R^2 \xi_{\rm CR} W_{\rm CR},
\label{eq:cr-heating}
\end{eqnarray}
where $n_{\rm ice}$ is the number density of ice molecules, $E_b$ is the binding energy of molecules making up the ice, and $\xi_{\rm CR}W_{\rm CR}$ is the energy flux of CRs in the Galactic disk. 

Above, we have calibrated approximately this energy flux using the dimensionless parameter $\xi_{\rm CR}$ and the local CR energy flux 
\begin{eqnarray}
W_{\rm CR}=\int^{E_{\rm max}}_{E_{\rm min}}4\pi j_{\rm LISM}(E)E\df E,
\label{eq:WCR}
\end{eqnarray}
where $E$ is the kinetic energy of CR particles and $j_{\rm LISM}(E)$ is the intensity of Galactic CR protons in the local ISM. 

For the local CR intensity, $\xi_{\rm CR}=1$, but its value is expected to change across the galaxy. Indeed, observations of diffuse gamma-ray emission reveal a higher density of GeV CRs than the local value for the inner 5 kpc from the Galactic Centre (\citealt{acero2016}, \citealt{yang2016}; see also \citealt{gabici2019} for a review) which agrees relatively well with models of Galactic CR transport \citep[see e.g.][]{recchia2016}. Recent analyses of gamma rays from giant molecular clouds (GMCs) also suggest a spatial variation of GeV CR density but instead for regions with galactocentric distances from 4 to 8 kpc \citep{aharonian2020}. In fact, the standard paradigm for Galactic CRs suggests that these particles are accelerated from discrete sources like supernova remnants (see e.g. \citealt{ahlers2009,morlino2012,gabici2014,mertsch2020,phan2020} and references therein). The discreteness of sources might lead CR density fluctuations even on smaller scales \citep{blasi2012,genolini2017,phan2021}. Even though the spatial variation of CR density is still uncertain, an enhancement of CR flux by a factor of a few might be important for the destruction of ISOs. For simplicity, we might take into account this effect by varying the values of $\xi_{\rm CR}$ (see Section \ref{sec:implications}). 

Assuming $V=4\pi R^3/3$ for a spherical ISO, the rate of change for the ISO radius could be estimated from Eq. \ref{eq:cr-heating} as follows
\begin{eqnarray}
\left(\frac{\df R}{\df t}\right)_{\rm CR}\simeq-\frac{\sigma_{\rm esc}}{4\pi R^2}\frac{\df V}{\df t}=-\sigma_{\rm esc}\frac{\xi_{\rm CR} W_{\rm CR}}{n_{\rm ice} E_b}
\label{eq:RCR}
\end{eqnarray}
where $\sigma_{\rm esc}$ represents the fraction of evaporated ice molecules that could escape through the CR tracks. Molecular dynamics simulations seem to suggest $\sigma_{\rm esc}\simeq 0.1$ for simple molecular solids \citep{bringa2002} and we shall adopt this value in the following. Recent studies of low-energy CRs by Voyager provide us with the spectrum down to kinetic energy of around a few MeVs \citep{cummings2016} and, thus, we shall choose $E_{\rm min}=1$ MeV. Note that using lower values for $E_{\rm min}$ would not alter the results significantly. In the following numerical estimate, we adopt the form of $j_{\rm LISM}(E)$ as provided in \citet{phan2018}, which has been obtained by fitting observed data from Voyager 1 \citep{cummings2016} and AMS-02 \citep{AMS2015}. From Eq.~\eqref{eq:WCR}, the corresponding local CR energy flux is $W_{\rm CR}\simeq 1.3\times 10^{10}$ eV cm$^{-2}$ s$^{-1}$. 

It is worth mentioning that the CR erosion rate obtained in this way is much larger than the previous estimate by \citet{jackson2021} mainly due to the different binding energy adopted in Eq.~\eqref{eq:cr-heating} and \eqref{eq:gas-heating}. The authors have adopted an effective binding energy $E_b^{\rm eff}=26$ eV for N$_2$ ice which is much larger than the typical binging energy of $E_b\simeq 0.07$ eV. In fact, the value of $E_b^{\rm eff}$ has been inferred from the experimental data showing the impact of low-energy ions (about 1 MeV per nucleon) on a small N$_2$ ice fragment of size roughly 20 $\mu$m \citep{vasconcelos2017}. However, it is not straightforward to extrapolate the results to the ISOs of interest as they are about \textit{one million times larger} in size. More importantly, the large difference in size also means that the CR energy range relevant for ISO heating is much larger than the one for small ice samples. For example, while ice mantles of micrometer-sized dust grains are heated mostly by MeV CRs \citep{ivlev2015}, the bulk of the deposited energy for ISOs are more likely from GeV CRs as they could form deep tracks inside these objects and lead to the transient evaporation explained above.

\subsection{Collisional Heating by Interstellar Gas}
Icy ISOs could also be destroyed as they collide with the ambient gas in the ISM. The rate of change in volume might be estimated by balancing the energy flux due to collisions and the one required to evaporate the ice \citep{hoang2020},
\begin{eqnarray}
n_{\rm ice}E_b\left(\frac{\df V}{\df t}\right)_{\rm gas}=\pi R^2 n_{\mathrm{H}}v_{\rm obj} \frac{1}{2}\mu m_{\mathrm{H}}v_{\rm obj}^2,
\label{eq:gas-heating}
\end{eqnarray}
where $n_{\text{H}}$ is the density of hydrogen atoms in the surroundings, $\mu\simeq1.4$ is the average atomic mass of the ISM gas, and $v_{\rm obj}$ is the ISO's speed. We could then obtain the evaporation rate as follows
\begin{eqnarray}
\left(\frac{\df R}{\df t}\right)_{\rm gas}\simeq -\frac{\sigma_{\rm esc}}{4\pi R^2}\left(\frac{\df V}{\df t}\right)_{\rm gas}=-\sigma_{\rm esc}\frac{n_{\text{H}}\mu m_{\text{H}}v_{\rm obj}^3}{8n_{\rm ice}E_b}.\n\\
\label{eq:Rgas}
\end{eqnarray}
% \begin{eqnarray}
% t_{\rm gas}\simeq\frac{V}{\df V/\df t}=\frac{8n_{\rm ice}E_bR}{3n_{\text{H}}\mu m_{\text{H}}v_{\rm obj}^3}.
% \label{eq:tgas}
% \end{eqnarray}
It is clear from Eq. \eqref{eq:Rgas} that heating due to gas collisions is more important in dense environments or for objects with relatively high speeds. Note that, since we are interested in the interstellar journey of ISOs, we shall restrict ourselves to the ISM gas density $n_{\rm H}\simeq 1$ cm$^{-3}$ in our numerical results. 

\subsection{Evolution of Icy Interstellar Objects}
We could now combine rate of change in radius due to both CRs and interstellar gas from Eq. \eqref{eq:RCR} and \eqref{eq:Rgas} for the total erosion rate as follows
\begin{eqnarray}
\frac{\df R}{\df t}=-\frac{\varepsilon\sigma_{\rm esc}}{n_{\rm ice}E_b}\left(\xi_{\rm CR}W_{\rm CR}+\frac{n_{\rm H}\mu m_{\rm H}v_{\rm obj}^3}{8}\right)
\label{eq:RISO}
\end{eqnarray}
where we have introduced also $\varepsilon$ as a surface-to-volume enhancement factor relative to a sphere. This allows us to approximately take into account arbitrary shapes of ISOs. For example, the analysis of \citet{mashchenko2019} provided the best-fit shape for 'Oumuamua to be either an oblate (pancake-shaped) ellipsoid, with axis ratios of 6:6:1; or a prolate (cigar-shaped) ellipsoid, with axis ratios 8:1:1. In both cases, the surface-to-volume enhancement factor could be estimated roughly as $\varepsilon\simeq s/2$ with $s$ being the ratio of major to minor axis, assuming $R$ in Eq.~\eqref{eq:RCR} and \eqref{eq:Rgas} corresponds to the major axis of the ellipsoid. We shall adopt $\varepsilon=3$ as in the oblate case for the numerical estimate.

We could now integrate both sides of Eq. \eqref{eq:RISO} to obtain the evolution of the ISO's radius in time
\begin{eqnarray}
%R(t)=R_0-\int_0^t\left[\left(\frac{\df R}{\df t'}\right)_{\rm CR}+\left(\frac{\df R}{\df t'}\right)_{\rm gas}\right]\df t'
R(t)=R_0-\frac{\varepsilon\sigma_{\rm esc}}{n_{\rm ice}E_b}\left(\xi_{\rm CR}W_{\rm CR}+\frac{n_{\rm H}\mu m_{\rm H}v_{\rm obj}^3}{8}\right)t,\n\\
\end{eqnarray}
where $R_0$ is the initial radius of the ISO. The destruction time of the ISO could be roughly estimated by setting $R(t=\tau)=0$, yielding 
\begin{eqnarray}
\tau=\frac{R_0n_{\rm ice}E_b}{\displaystyle\varepsilon\sigma_{\rm esc}\left(\xi_{\rm CR}W_{\rm CR}+\frac{n_{\rm H}\mu m_{\rm H}v_{\rm obj}^3}{8}\right)}.
\end{eqnarray}

\section{Ice Modification\\ for Interstellar Objects}
\label{sec:modification}
Having derived the evolution of the ISO's radius in time, we could also discuss the ice modification for ISOs due to CR heating. As mentioned in Sec. \ref{sec:destruction}, CRs could create hot tracks of evaporated molecules inside ISOs and only a small fraction of these molecules could escape into the ISM. The molecules remaining inside the tracks might quickly cool down to the surrounding ice temperature and close the track \citep{mainitz2016}. However, we expect the ice structure to be different from before being heated by CRs and the rate of change for the modified volume $\df V_{\rm m}/\df t$ could be roughly estimated also from energy conservation as in Eq. \eqref{eq:cr-heating} and Eq. \eqref{eq:gas-heating}
\begin{eqnarray}
\frac{\df V_{\rm m}}{\df t}&=&-(1-\sigma_{\rm esc})\left(1-\frac{V_m}{V}\right)\frac{4\pi R^2}{\sigma_{\rm esc}}\frac{\df R}{\df t}
% \\
% \Rightarrow &&\frac{\df V}{\df t}=(1-\sigma_{\rm esc})\left(1-\frac{V_m}{V}\right)\frac{\df V}{\df t}\n\\,
% \Rightarrow && V^{1-\sigma_{\rm esc}}\frac{\df }{\df V}\left[V^{-(1-\sigma_{\rm esc})}V_m\right]=-(1-\sigma_{\rm esc})
\label{eq:dVm}
\end{eqnarray}
where the factor $1-\sigma_{\rm esc}$ represents the fraction of particles remaining in the tracks and the factor $1-V_m/V$ takes into account the probability that the newly modified volume will not coincide with the previously modified volume. We could now adopt $R(t)$ from Eq. \eqref{eq:RISO} and solve Eq. \eqref{eq:dVm} to obtain the evolution of the filling factor $f_m=V_m/V$ in time
\begin{eqnarray}
f_{m}=
\left\{ \begin{array}{ll}
\frac{1-\sigma_{\rm esc}}{\sigma_{\rm esc}}\left\{\left[\frac{R_0}{R(t)}\right]^{3\sigma_{\rm esc}}-1\right\} & \quad t\leq\tau_m\\
1 & \quad t>\tau_m 
\end{array}
\right.\ ,
\end{eqnarray}
where $\tau_m=\left[1-(1-\sigma_{\rm esc})^{\frac{1}{3\sigma_{\rm esc}}}\right]\tau$ is the modification time at which the ISO is completely modified. Interestingly, $\tau_m<\tau$ for $\sigma_{\rm esc}<1$ meaning the ice structure would always be modified entirely before CR and gas heating could totally destroy the ISO. 

\section{Results for Interstellar Objects and Implications for `Oumuamua}
\label{sec:implications}

\begin{figure}[ht!]
\includegraphics[width=3.5in]{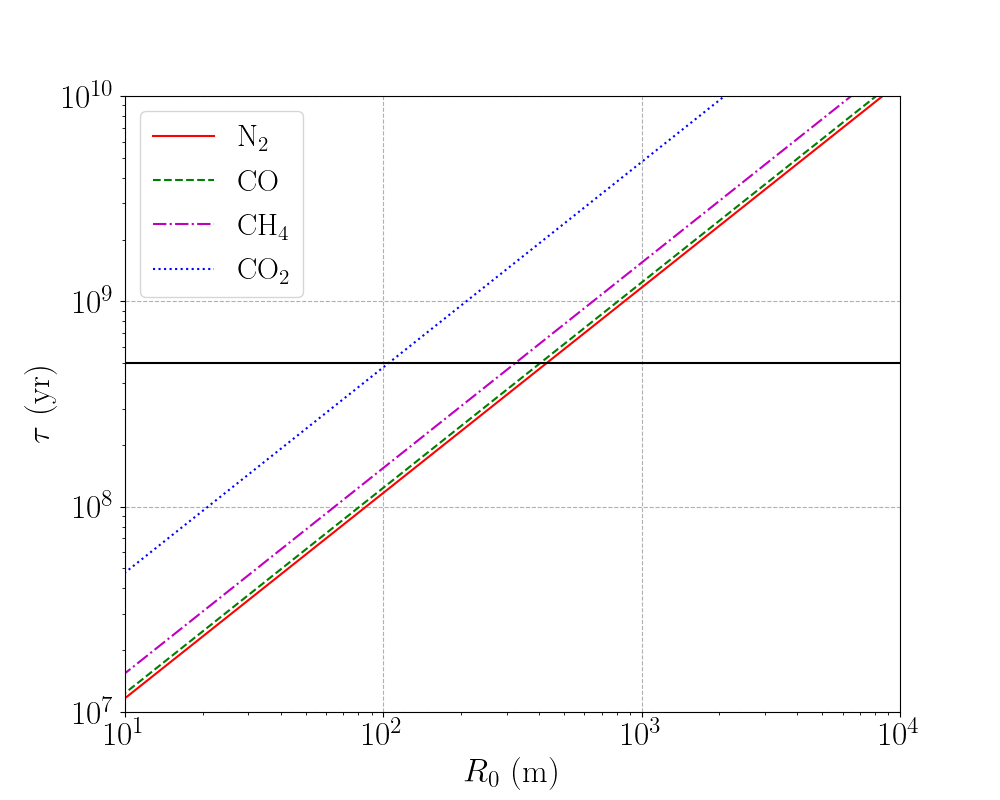}
\caption{Erosion time for various types of ices including N$_2$ (solid red line), CO (dashed green line), CO$_2$ (dotted blue line), and CH$_4$ (dash-dot magenta line) given $\xi_{\rm CR}=1$ and $\varepsilon=3$ in comparison with the suggested travel time of around 0.5 Gyr for `Oumuamua (solid black line).}
\label{fg:tau-R}
\end{figure}

The erosion times for different types of ices including N$_2$, CO, CO$_2$, and CH$_4$ with the ice densities and binding energies collected from \citet{jackson2021} are presented in Fig. \ref{fg:tau-R} for an object with speed $v_{\rm obj}\simeq 10$ km/s. We note that these results have been obtained using $\xi_{\rm CR}=1$ and the erosion time could be much shorter for ISOs originated from regions with an enhanced CR flux; e.g. $\xi_{\rm CR}\simeq 5$ has been inferred from gamma-ray data of GMCs in certain regions \citep{aharonian2020}.

Now let's consider an ISO created at a distance $D$ from the Solar System. The minimum initial size of this object could be constrained by balancing the traveling time $t_{\rm trav}=D/v_{\rm obj}$ with the destruction time, yielding
\begin{eqnarray}
R_{0,\rm min}=\frac{\varepsilon\sigma_{\rm esc}D}{n_{\rm ice}E_bv_{\rm obj}}\left(\xi_{\rm CR}W_{\rm CR}+\frac{n_{\rm H}\mu m_{\rm H}v_{\rm obj}^3}{8}\right). \label{eq:rmin} \n\\
\end{eqnarray}
Equation \eqref{eq:rmin} might have interesting implications for `Oumumua as \citet{jackson2021} had suggested that it could be an N$_2$ fragment originating from the Perseus arm about 0.5 Gyr ago. This means that the initial size of `Oumuamua should be between about $0.5$ to $2.5$ km, depending on the value of $\xi_{\rm CR}$, in order for it to survive the interstellar journey (see also Fig. \ref{fg:tau-R}).  

\begin{figure}[ht!]
\includegraphics[width=3.5in]{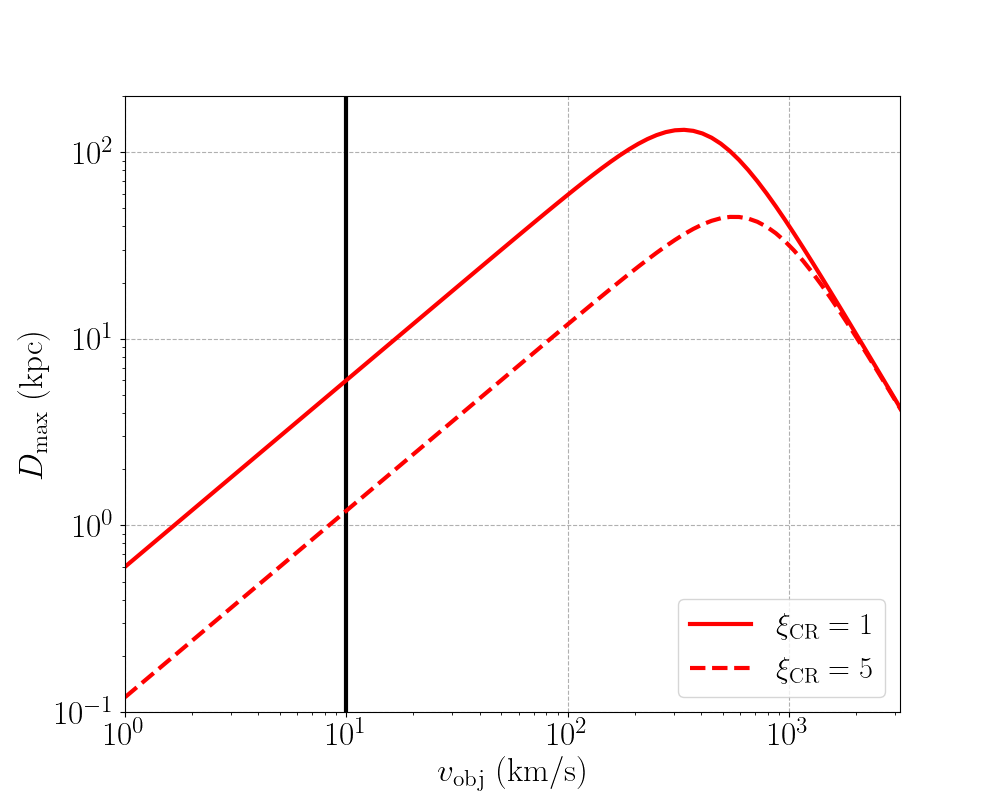}
\caption{Maximum distance to the birth site versus speed in the case where $\xi_{\rm CR}$=1 (red solid curve) and $\xi_{\rm CR}$=5 (red dashed curve) for an N$_2$ ice fragment with $R_0=0.5$ km and $\varepsilon=3$. The black vertical line marks $v_{\rm obj}= 10$ km/s comparable to the speed of `Oumuamua (see text for more details).}
\label{fg:Dmax}
\end{figure}

We could also turn the argument around and provide constraints on the distance to the birth site of an ISO for given values of $R_0$ and $v_{\rm obj}$. For an object to survive the interstellar journey, we expect $R_0\gtrsim R_{0,\rm min}$ which, from Eq.~\eqref{eq:rmin}, could be re-written as a condition for $D$
\begin{eqnarray}
D\lesssim D_{\rm max}=\frac{R_0n_{\rm ice}E_bv_{\rm obj}}{\varepsilon\sigma_{\rm esc}\left(\xi_{\rm CR}W_{\rm CR}+\displaystyle\frac{n_\mathrm{H}\mu m_\mathrm{H}v_{\rm obj}^3}{8}\right)}.
\label{eq:Dmax}
\end{eqnarray}
If the initial size $R_0$ of an object is known presumably from a particular formation mechanism of ISOs, Eq.~\eqref{eq:Dmax} allows us to set the limit of the maximum distance $D_{\rm max}$ for different values of the object's speed $v_{\rm obj}$. Since collisions with ISM gas are the dominant erosion mechanism for high-speed ISOs, we expect $D_{\rm max}$ to be independent of the CR density above a characteristic speed, $v_{\rm c,obj}=\left[8\xi_{\rm CR}W_{\rm CR}/(n_{\rm H}\mu m_{\rm H})\right]^{1/3}$. Objects moving too slowly, on the other hand, would be destroyed by CRs before they could reach the Solar System and, thus, $D_{\rm max}$ for $v_{\rm obj}<v_{\rm c,obj}$ should be sensitive to the value of $\xi_{\rm CR}$. This is illustrated in Fig.~\ref{fg:Dmax} where we present $D_{\rm max}$ for $\xi_{\rm CR}$=1 (red solid curve) and $\xi_{\rm CR}$=5 (red dashed curve) given an N$_2$ ice fragment with $R_0=0.5$ km and $\varepsilon=3$ for different values of $v_{\rm obj}$. At a speed of $v_{\rm obj}\simeq 10$ km/s as in the case of `Oumuamua \citep{mamajek2017,meech2017}, $D\gtrsim D_{\rm max}\simeq 5 \textrm{ kpc}$ ($D\gtrsim D_{\rm max}\simeq 1 \textrm{ kpc}$) for a ten-kilometer N$_2$ ice fragment to reach the Earth for $\xi_{\rm CR}=1$ ($\xi_{\rm CR}=5$). It is also clear from this example that a more detailed study of the spatial profile of Galactic CRs might help to shed light on the origin of ISOs passing through the Solar System.   

\begin{figure}[ht!]
\includegraphics[width=3.5in]{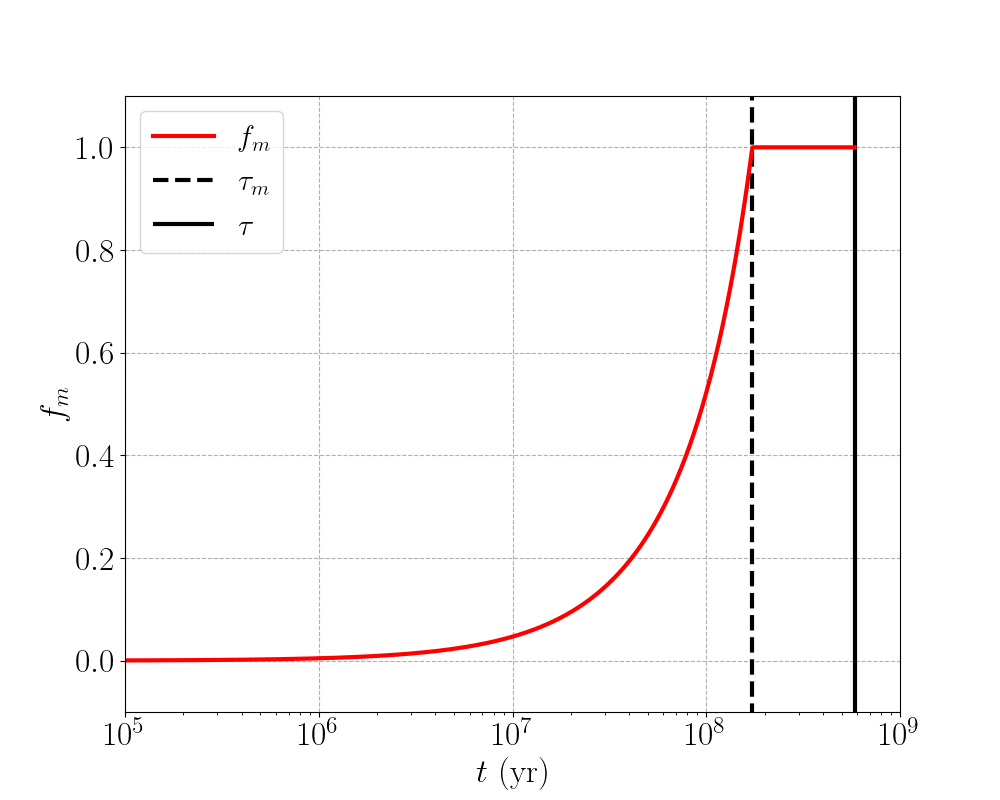}
\caption{Volume filling factor versus travel time for $\xi_{\rm CR}$=1 (red solid curve) for an N$_2$ ice fragment with $R_0=0.5$ km and $\varepsilon=3$. The black vertical dashed and solid lines mark respectively $\tau_m$ and $\tau$ which are the modification and destruction times (see text for more details).}
\label{fg:fm}
\end{figure}

We examine also the evolution of the volume filling factor in time for an N$_2$ ice fragment with $R_0=0.5$ km and $\varepsilon=3$ moving with $v_{\rm obj}=10$ km/s. The modification and destruction times for this ISO should be around $\tau_m\simeq0.17$ Gyr and $\tau=0.58$ Gyr which are marked as the dashed and solid vertical black lines in Fig. \ref{fg:fm}. The modification of the ice structure might have interesting observational consequences as it is well known that the spectroscopic features might be different for amorphous or crystal ice structures. Again if the volume filling factor of the amorphous ice structure could also be assumed from a formation mechanism, observational constraints could be made for the origin of the ISOs.

\section{Conclusions}
\label{sec:conclusions}

We have studied the destruction and modification of icy ISOs and found that ISOs could be completely modified before being destroyed by CRs and interstellar gas. This might have interesting implications for spectroscopic studies of ISOs for future flyby/intercept missions \citep{heina2020,moore2021}. We also identified conditions for these objects to survive their interstellar journey as follows: 
\begin{enumerate}
\item Cosmic-ray heating and ISM gas collisions are both important erosion mechanisms for icy ISOs. In fact, the destruction rate of CRs might be more efficient than previously suggested by \citet{jackson2021} such that the initial size of an N$_2$ fragment involved to explain many peculiar properties of `Oumuamua should be larger than $R_{0,\rm min}\simeq 0.5$ km for the surface-to-volume enhancement factor $\varepsilon=3$ if its travel time in the ISM is about 0.5 Gyr as proposed by these authors. This exacerbates the mass budget problem of this scenario \citep{siraj2021,levine2021}. One might expect an even larger value for $R_{\rm min}$ in the case where the ice fragment originates from a region with a higher CR density than locally observed.

\item For ISOs created at a distance $D=5$ kpc with $\varepsilon=3$ and $v_{\rm obj}\simeq 10$ km/s, their minimum initial sizes should be between $R_{0,\rm min}\simeq 0.1$ to $0.5$ km for common types of ices such as N$_2$, CO, CO$_2$, and CH$_4$ assuming the CR flux similar to the local one and $\sigma_{\rm esc}=0.1$. This represents an increase by at least an order of magnitude in the required initial N$_2$ mass of `Oumuamua relative to its final value. If the CR density is higher by a factor $\xi_{\rm CR}$ which could be between 2 to 5 in certain regions \citep{aharonian2020}, the values of $R_{0,\rm min}$ should be scaled accordingly. 

\item The relatively short erosion time due to CR heating and gas collisions also allow us to limit the potential range of the maximum distance to the birth site for ISOs given the initial size $R_0$ and the speed $v_{\rm obj}$. For example, if `Oumuamua is composed of N$_2$ ice, it should have formed within $D_{\rm max}=4$ kpc (or $D_{\rm max}=1$ kpc) from the Solar System given $v_{\rm obj}\simeq 10$ km/s, $R_0=0.5$ km and $\xi_{\rm CR}\simeq 1$ (or $\xi_{\rm CR}\simeq 5$). It would be interesting to incorporate a detailed modeling of the CR distribution in the Galactic disk to set more rigorous constraints on the birth site of known ISOs and this might help to provide a more stringent upper limit for $D$ and better clarify their origin.  
\end{enumerate}

We would like to thank the organizers of VLLT Joint Seminar Series, especially Le Ngoc Tram, Chi Thanh Nguyen, Hoang-Dai-Nghia Nguyen, and Nhat-Minh Nguyen. The series has brought about many fruitful discussions which started this project. V.H.M.P. would like to thank Philipp Mertsch and Stefano Gabici for reading the manuscript and provding helpful comments. T.H. acknowledges the support by the National Research Foundation of Korea (NRF) grant funded by the Korea government (2019R1A2C1087045). A.L. was supported in part by a grant from the Breakthrough Prize Foundation.

\bibliography{mybib}
\bibliographystyle{aasjournal}

\end{document}